# *Insider threats in Cyber Security: The enemy within the gates.*

*Guerrino Mazzarolo, Anca Delia Jurcut*

*School of Computer Science, University College Dublin, Ireland*

*{guerrino.mazzarolo\*, anca.jurcut\*} @ucd.ie*

## *Introduction:*

The Capital One data leakage - as the latest instance of exploitation carried out by the actions of an insider threat - "involves the theft of more than 100 million customer records, 140,000 Social Security numbers and 80,000 linked bank details of Capital One customers, allegedly stolen by a single insider, according to court filings in Seattle" (Kate Fazzini, CNBC). Cybercrime continues to exhibit rising trends. According to a study provided by Ponemon Institute and Accenture, there was an 11% increase in the average annual number of security breaches in 2018 (Ponemon Institute LLC and Accenture, 2019). A data breach is an incident in which protected data has been accessed or disclosed in an unauthorized fashion. Those kinds of incidents can be caused by internal or external actors. Internal ones are either malicious or careless users. The external threat category includes hackers, cybercriminal organizations and state-sponsored actors. The data breaches that appear on the news are typically carried out by outsiders. Attacks coming from the outside generally expose threats that have been addressed with traditional security measures through a "defense in-depth" approach. The hazards that originate from inside are more difficult to prevent and detect because insiders pose a high danger as they are familiar with the organization's network topology, systems, directives, and policies, and they have access to confidential information with relatively low restrictions. When we analyze cybercrime, often we underestimate the dangers of the internal threat. Insiders present a significant risk to organizations and, even if they were not the most common source of attacks in past years, they were the most expensive and difficult to recover from.

## *Definition*

There are two main types of insiders: malicious users (those that intentionally harm the institutions, as described above), and unintentional insider users (those that accidentally expose confidential data, as described below). These activities endanger confidentiality, integrity and availability of the business.

Different definitions of malicious insider threat could be found. The one from CERT US provides a comprehensive explanation. "A malicious insider threat, may be either a current or former employee, contractor, or business partner who has or had authorized access to an organization's network, system, or data and *intentionally* exceeded or misused that access in a manner that negatively affected the confidentiality, integrity, or availability of the organization's information or information systems" (The Cert Guide to Insider Threats, 2012). Motivators behind malicious insiders could include: monetary gain, a disgruntled employee, entitlement, ideology, or outside influence with the consequences of fraud, sabotage, espionage and theft or loss of confidential information.

The official working definition from the report "Unintentional Insider Threats: A Foundational Study" states that an *unintentional* insider threat is: "a current or former employee, contractor, or business partner who has or had authorized access to an organization's network, system, or data and who, through action or inaction *without malicious intent*, causes harm or substantially increases the probability of future serious harm to the confidentiality, integrity, or availability of the organization's information or information systems" (The Cert Guide to Insider Threats, 2013).

Insider threats could be considered the biggest cybersecurity danger to firms, organizations and government agencies. According to the security company Clearswift, "Organizations report that 42% of IT security incidents occur as a result of their employees' actions" (Clearswift, 2017).

*Current Concerns*

The dangers posed by insiders have multiple variables that must be considered when the risk is assessed. The following list provides an outline of the most important tasks that need to be addressed in order to be aware of the hazard introduced by internal actors.

Firstly, an insider can easily bypass existing physical and technological security controls through legitimate rights. Employees, in fact, need to access the organization's data for their daily tasks. Spotting the malicious activity is extremely difficult and time consuming. In addition, members of staff with technical knowledge can elude security controls currently in place.

Another factor that needs to be considered is that technology alone is not enough. Controls can detect and block malevolent actions; however, if we want to prevent indiscriminate users, we need additional information. Human behaviour has a huge influence in comportment and sometimes can provide indications that help to identify a person's further actions. Emerging techniques are focusing on sentiment analysis. Monitoring and identifying disgruntled employees could greatly increase chances of isolating promiscuous activity.

In order to prevent malevolent activities, we need data from different controls. The availability of this information is often jeopardized within an entire organization and by being available to a different business owner. The difficulty is to have the approval of different stakeholders in order to have legal access to the data. Once these are aggregated, it is important to add correlation rules to the raw data so it can give us significant information to help our Security Operation Center (SOC) to be alerted in case of any infraction or suspicious behaviour.

Safeguarding the privacy of the data in accordance with the law is mandatory for every firm and organisation. It is necessary that insider threat analysts follow laws regarding privacy, civil liberties, and legal guidance, including the General Data Protection Regulation (Regulation (EU) 2016/679 of the European Parliament and of the Council, 2016).

Preventing data leakage does not stop at the boundary of a firm's network and it goes further beyond. The continued interaction with different business partners exposes systems to another threat. It is difficult to control the cyber security standards of affiliates. If not monitored properly, dishonest employees could use this as an entry point to an organization's data and system, and misuse it.

*Types of Insider Activities*

A study conducted by NATO's Cooperative Cyber Defense Centre of Excellence (Kont, Pihelgas, Wojtkowiak, Trinberg, and Osula, 2015) categorizes threats into five mains different areas:

1. Fraud - consisting of the use of the company's information and data for personal gain.

2. IT sabotage – which is a major and unpredictable action against the firm and could seriously impact the availability of the overall infrastructure.

3. Intellectual property theft – which is a remunerative action that permits insiders to exfiltrate copyrights, patents, trademarks, and trade secrets without permission.

4. Espionage - the practice of illegally obtaining information about the plans and activities from industrial or international government entities.

5. Unintentional – employee that have not malevolent intent, although their actions, or behavior, occasionally affect the organization.

Unintentional actors, as a result of carelessness, also can lead to a major security breach. This could cause as much damage as malevolent actions. Disclosure of classified information in the public domain and social media, accidentally replying to phishing campaigns, or downloading malicious code off the Internet are all examples of negligent activities that could have serious consequences for the organisation. A famous example within this category was the breach perpetrated on the American security company RSA in 2011. Four employees were targeted through a phishing email campaign, one of them clicked on the attachment that used a zero-day exploit, targeting a vulnerability. The intruders succeeded in exfiltrating confidential information related to the company's SecurID two-factor authentication products (Zetter, 2011).

As technology grows and develops also malicious actors are constantly evolving and impacting our society. "Hacktivist" is a new threat category that was reported recently on the breaking news. The word "hacktivism" is relatively novel in the cyber domain. It combines the word "hacker" and "activist". It consists of gaining unauthorized access to ICT systems and carrying out a variety of disruptive actions as a means of achieving political and social goals. In cases such as those of Chelsea Manning and Edward Snowden, thousands of classified documents were leaked, and a large amount of confidential information was revealed. Those incidents involve a number of distinct forms of cybercrimes: sabotage, espionage, intellectual property theft, and mark the difficulties of preventing a malicious actor from dispersing data even if sophisticated controls are in place (Ewen MacAskill, 2017).

*Proposed Remediation Guidelines*

To increase the chances of preventing, detecting and responding to insider threats, the overall security of the organisation must be well structured and organised.

Every project needs executive-level managers leadership. In a company, a chief information security officer (CISO) must provide a long-term vision, engage with all departments, and finally promote and build a strong insider threat program (InTP). An effective InTP must include participation from different stakeholders. Mandatory departments that could contribute actively to the project are human resources, information assurance, legal, and cyber security.

Best practices for an InTP should take into consideration at least the following areas in order to decrease the overall risk: administrative controls, technical controls, physical controls, security awareness and incident response.

1- Administrative controls such as policies, directives and regulations must be clearly documented and enforced. It is important to show the acceptable use of an organisation's system, network and information. It is relevant to remark what is expected from employees and also the possible consequences of violations. This is a valuable deterrence method.

2 - Technical control is generally considered the backbone of every InTP. Data loss prevention, email monitoring, web proxy, rogue device detection, endpoint analytics, security information and event management (SIEM) are safeguards that can detect and prevent data leakage. In recent years, security has developed a new capability called User and Entity Behaviour Analytics (UEBA). This includes a behavioural analysis of entities, other than the users, such as routers, servers and endpoints. UEBA is much more effective since it can analyze the behaviour across multiple users and ICT devices in order to detect complex attacks.

3 - Physical controls are also well-founded in assigning the least privilege i.e. only enough access to perform the required job and for implementing a segregation of duty when more than one person is required to complete a critical function. Physical controls are also used to control and minimise the risk of unauthorised access to physical assets and information systems.

4 - The employees in a company need to have security awareness training, with A specific chapter dedicated to insider threat in order to explain what is expected from them and, which threats they might be exposed to. Unintentional insider threat may be recruited from outside through the social engineering. It is important to make all employees acknowledgeable about such hazards and to train them how to avoid to be the weakness in the organisation's security barrier.

5 - Finally, the organisations must be ready to respond to an incident involving a malicious or unintentional insider threat. The incident response (IR) workflow should be part of an existing plan. However, the process should also include the escalation of the reporting, the notification to management, and submission to an investigation officer.

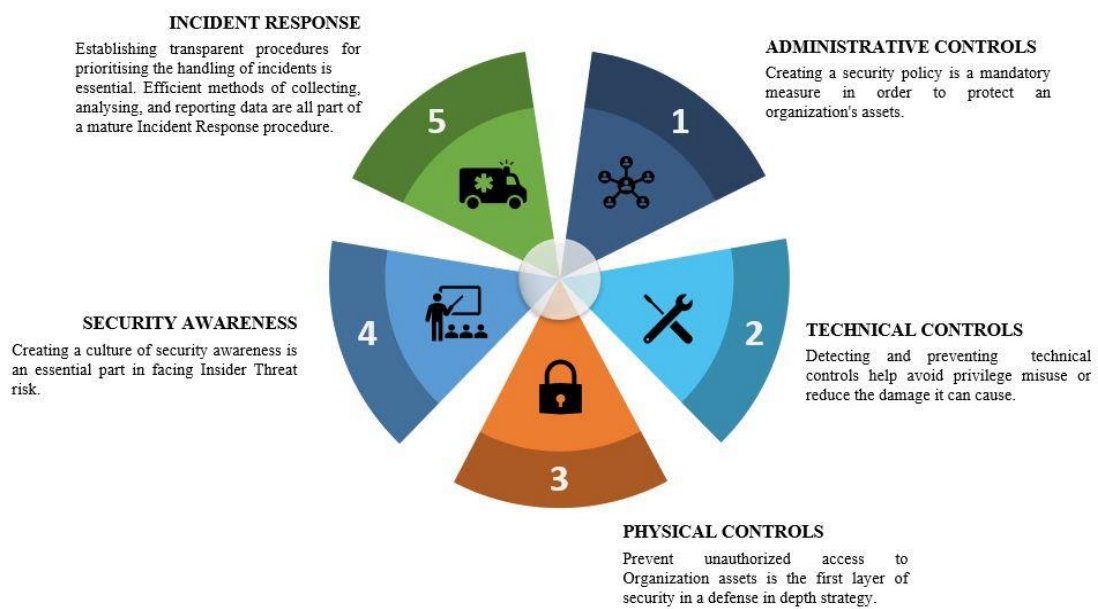

Figure 1. Proposed Remediation Guidelines.

*Future Challenges*

A new challenge appeared in the recent years. This concerns the adoption of cloud and mobile technology in companies, and it transformed the IT infrastructures considerably. Physical boundaries of corporate networks and digital assets become less clearly defined than it used to be in the past. Therefore, new challenges are calling for new approaches.

Since it is difficult to stop an insider threat at the boundary, early detection is the key. In modern ICT infrastructure, several technical controls are implemented and could identify suspicious activities such as unauthorised access, violation of organisation policies, internal reconnaissance, abuse of rights, and data loss. However, cyber security should not concern just technology. This approach can prevent and control the possible damage, but nowadays this approach it is not enough. It should be integrated with an additional layer. Guarding information and systems against insiders with illegitimate intentions requires a multidimensional defensive strategy.

Analyzing user personalities through social media and combining the information with the technical domain could represent a dynamic approach to decrease the likelihood of such threats. Marwan Omar remarks: "Organizations need to implement multi layered defensive approaches to combat insider risks" (Marwan, 2015, p. 162).

Nowadays, technical controls are already present in most organisations' security programs. However, because of an increasingly threatening landscape, they cannot be solely relied upon. New technologies are periodically proposed, which potentially offer to the malicious employees new opportunities to strike. Using innovative know-how as a countermeasure should be the main priority.

Technologies based on machine learning and artificial intelligence are to be implemented in order to assist with prevention and detection of insider threats before they are causing irreversible damage. The future development chapter of Cyber security has yet to be written, particularly with the coming power of quantum computers.

### *Conclusion:*

Insider threat employees represent the wolf in sheep's clothing. Daily, real examples, clearly show that insider threats create a significant hazard to every company, institution and organisation. A potential malicious insider can cause millions of euro in damage by stealing intellectual property, sabotaging facilities or disclosing information that can irreparably compromise the organisation. However, an unintentional insider can cause irreversible damage as well.

Enterprises will never be able to fully make sure that employees have no malicious intentions, or that they won't ever fall for phishing email campaigns. Meanwhile, although the elimination of all risks is not possible, the overall risks could be reduced and the residual risk controlled. Too often, the security strategy is dedicated to the edge security layer and ignores that a conspicuous threat might come within the organization.

To conclude, defending your enterprise from insider threats is a vital part of information security best-practices. It is essential that your company's highly valuable classified data and assets are protected from its greatest threat: the enemy within the gates. Every mature cyber security program nowadays should contain an insider threat assessment and a comprehensive insider threat program to protect a corporation's people, facilities, networks, and intellectual property.